\pdfoutput=1
\documentclass[aps,pra,twocolumn,superscriptaddress]{revtex4}

\usepackage{amssymb}
\usepackage{amsmath}
\usepackage{amsthm}
\usepackage{graphicx}
\usepackage{verbatim}
\usepackage[normalem]{ulem}
\usepackage{color}
\usepackage{url}
\usepackage{hyperref}

\begin{document}

\title{One-dimensional Bose gas in optical lattices of arbitrary strength}

\author{Grigory E. Astrakharchik}
\affiliation{Departament de F\'{\i}sica, Universitat Polit\`{e}cnica de Catalunya, 08034 Barcelona, Spain}

\author{Konstantin V. Krutitsky}
\affiliation{Fakult\"at f\"ur Physik der Universit\"at Duisburg-Essen Campus Duisburg, 47048 Duisburg, Germany }

\author{Maciej Lewenstein}
\affiliation{ICFO – Institut de Ci\`{e}ncies Fot\`{o}niques, The Barcelona Institute of Science and Technology, 08860 Castelldefels, Spain}
\affiliation{ICREA--Instituci\'o Catalana de Recerca i Estudis Avan\c{c}ats, Lluis Companys 23, 08010 Barcelona, Spain}

\author{Ferran Mazzanti}
\affiliation{Departament de F\'{\i}sica i Enginyeria Nuclear, Universitat Polit\`{e}cnica de Catalunya, Barcelona, Spain}

\date{February 8, 2016}


\begin{abstract}
One-dimensional Bose gas with contact interaction in optical lattices at zero temperature is investigated by means of the exact diffusion Monte Carlo algorithm.
The results obtained from the fundamental continuous model are compared with those obtained from the lattice (discrete) Bose--Hubbard model, using exact diagonalization, and from the quantum sine--Gordon model.
We map out the complete phase diagram of the continuous model and determine the regions of applicability of the Bose--Hubbard model.
Various physical quantities characterizing the systems are calculated and it is demonstrated that the sine--Gordon model used for shallow lattices is inaccurate.
\end{abstract}

\maketitle

The Bose--Hubbard model (BHM) was introduced in 1963~\cite{Gersch63,Hubbard1963}.
While the original motivation was to describe a crystalline solid, for which the model failed, the BHM became one of the fundamental quantum many-body problems.
It has found clear-cut realization with ultracold atoms in deep optical lattices.
This lead to the seminal observation~\cite{Greiner02} of the superfluid--Mott-insulator quantum phase transition~\cite{Fisher1989}
following the proposal of Ref.~\cite{Jaksch1998}.
In many aspects the experiments surpass the theory as shallow optical lattices can be easily realized, while no exact quantum many-body description of such systems is known up to date.
Even the case of deep optical lattices is controversial, as the scattered discussions demonstrate~\cite{Zwerger03,Lewenstein07,Bloch08,Cazalilla11,Lewenstein12,KrutitskyROPP} indicating the necessity to go beyond the standard BHM (for a review see~\cite{DuttaROPP}).
Nevertheless, the BHM is commonly used for lattice systems in different dimensions and it frequently works very well.
Still, there arise natural and important questions that motivated the present work: When can it be used with confidence? What is the regime of validity of the BHM?

The discrete BHM is derived from a continuous space model that, due to its complexity, has only been addressed  recently~\cite{SCG14,NHTP14,SG12,SDS10,Sakhel12,Pilati12}.
In this Letter, we use the {\em exact} diffusion quantum Monte Carlo method~\cite{Mazzanti08,SG12,Carbonell2013,SCG14,CSG14} and investigate one-dimensional Bose gas in optical lattices using a continuous Hamiltonian in real space.
We compare the results with those obtained from the BHM and determine its regions of validity.
Furthermore, a whole new generation of clean experiments on one-dimensional Bose gases loaded in optical lattices~\cite{Paredes04,Kinoshita04} have appeared, while the comparison of theory with experiment is not perfect~\cite{Haller10,Haller11}.
We also analyze the sine--Gordon (SG) model, commonly used for shallow lattices~\cite{Giamarchi88,Giamarchi04,Cazalilla11}, and show that it cannot be straightforwardly used for predicting for the position of the phase transition and the value of the gap~\cite{Haller10}.
We calculate the static structure factor, the one-body density matrix, the energy gap, the Luttinger parameter and its dependence on the interaction and lattice strengths.
Finally, we compare our results with the experiments of Ref.~\cite{Haller10} -- surprisingly our theory, which is in principle superior to all approximate ones, does not always provide a better description.

The first quantization Hamiltonian of $N$ bosons of mass $m$ interacting by a contact potential of strength $g=-2\hbar^2/(m a_{\rm 1D})$,
with $a_{\rm 1D}$ being the one-dimensional $s$-wave scattering length, has the form
\begin{equation}
\hat H
=
\sum_{i=1}^N
\left[
    -{\hbar^2 \over 2m}
    {\partial^2 \over \partial x_i^2}
    +
    V_{\rm L}(x_i)
\right]
+
g
\sum_{i<j}
\delta(x_i-x_j)
\;.
\label{Eq:H:continuous}
\end{equation}
The external potential $V_{\rm L}(x) = V_0\cos^2(\pi x/a_0)$ represents an optical lattice of strength $V_0$ with the lattice constant $a_0$.
A characteristic energy associated with the lattice is the recoil energy $E_{\rm rec} = \pi^2\hbar^2/(2ma_0^2)$.
We consider a system of finite size $La_0$, where $L$ is an integer, and impose periodic boundary conditions.

The ground-state properties of Hamiltonian~(\ref{Eq:H:continuous}) are studied using the Diffusion Monte Carlo (DMC) algorithm~\cite{Casu95} that solves the Schr\"odinger equation in imaginary time.
Statistical variance is significantly diminished by using the importance sampling.
The DMC method gives exact estimation of any observable commuting with the Hamiltonian, and delivers bias-free predictions for other observables by pure estimator techniques~\cite{Casu95}.

In deep optical lattices, model Eq.~(\ref{Eq:H:continuous}) reduces to the BHM.
In its standard and simplest form, the second quantization Hamiltonian is given by
\begin{equation}
\hat H_{\rm BH}
=
-J
\sum_{\ell=1}^L
\left(
    \hat a_\ell^\dagger
    \hat a_{\ell+1}^{\phantom{\dagger}}
    +
    {\rm h.c.}
\right)
+
\frac{U}{2}
\sum_{\ell=1}^L
\hat a_\ell^\dagger
\hat a_\ell^\dagger
\hat a_\ell^{\phantom{\dagger}}
\hat a_\ell^{\phantom{\dagger}}
\label{Eq:H:BH}
\end{equation}
The hopping and interaction constants $J$ and $U$ are determined as~\cite{Lewenstein12}
\begin{eqnarray}
\label{Eq:JU}
J
&=&
-\int_{0}^{La_0}
W_{\ell}^*(x)
\left[
    -
    \frac{\hbar^2}{2m}
    \frac{\partial^2}{\partial x^2}
    +
    V_{\rm L}(x)
\right]
W_{\ell+1}(x)
\,dx
\;,
\nonumber\\
U
&=&
g
\int_{0}^{La_0}
\left|
    W_{\ell}(x)
\right|^4
\,dx
\;,
\end{eqnarray}
where $W_{\ell}(x)$ is the Wannier function for the lowest Bloch band (maximally) localized near the minimum $x=x_\ell$
of the periodic potential $V_{\rm L}(x)$.
The results for the BHM are obtained by exact diagonalization.

\begin{figure}[t!]
\includegraphics[width=0.8\columnwidth, angle=0]{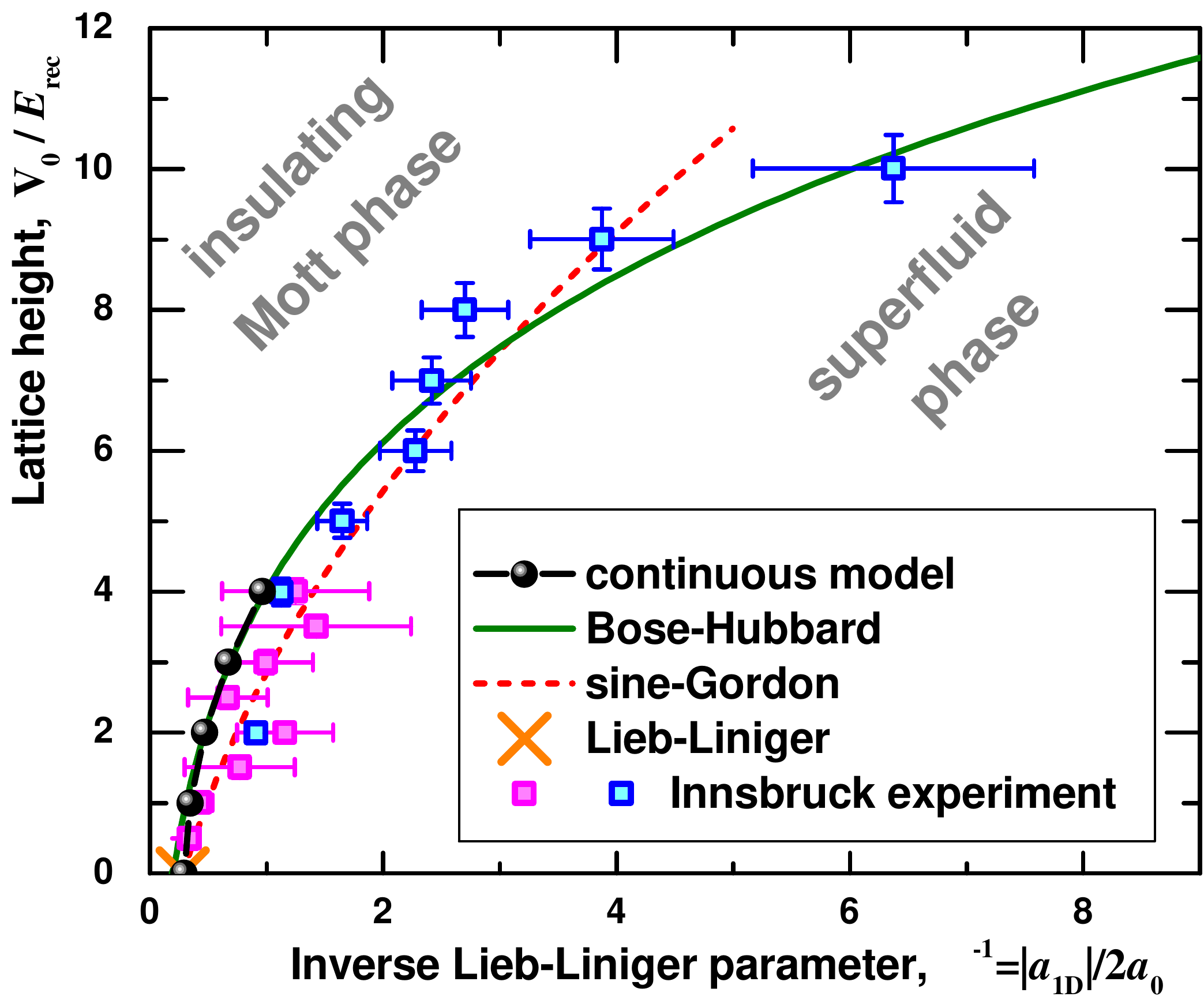}
\caption{(Color online)
The zero-temperature phase diagram of the continuous model compared with various approximate theories, as a function of the $s$-wave scattering length $a_{\rm 1D} / a_0$ and the optical lattice intensity $V_0 / E_{\rm rec}$.
The position of the Mott-insulator--superfluid phase transition is defined by $K=2$.
Black circles, DMC results;
green solid line, exact diagonalization of the BHM;
dashed red line, sine--Gordon predictions;
squares with error bars,
(pink - amplitude modulation,
blue - transport measurements)
experimental results of Ref.~\cite{Haller10}.
DMC and exact diagonalization results are obtained for $N=12$.
} \label{figure1}
\label{FigEN1}
\end{figure}

Figure~\ref{figure1} presents the complete phase diagram of the continuous model compared with various theories.
The transition line separating the superfluid and the Mott insulator phases is obtained from the Luttinger parameter $K = v_\text{F}/c$, where the Fermi velocity $v_\text{F} =\hbar\pi N/(L a_0 m)$ is entirely fixed by the system setup, while the speed of sound $c$ depends in a non-trivial way on the strength of the interaction and the lattice height~\cite{Giamarchi04,Cazalilla11}.
For unit filling, the phase transition takes place at the critical value $K=2$ as it follows from effective renormalization group theory~\cite{Giamarchi04}.
The phase transition of the continuous model is shown by black circles.
For $V_0=0$ the critical value $|a_{\rm 1D}|/a_0 = 0.56$ coincides with that of the Lieb--Liniger model.
It is interesting to compare with the SG and BHM, which are expected to be valid for shallow lattices and high lattices with weak interactions, respectively.
As there is no way to establish the exact regions of applicability of each theory internally, we deduce them by direct comparison with the DMC data.
Within the BHM, the transition is governed by a single parameter, $J/U$.
Using the $K=2$ criteria for $N=12$ particles one obtains a critical point at $(J/U)_{\rm c}=0.28$.
The relation to the two parameters of the continuous model (the lattice intensity and the interaction strength) is obtained from Eqs.~(\ref{Eq:JU}), resulting in the solid green line in Fig.~\ref{figure1}.
We find that for $V_0/E_{\rm rec}\gtrsim 3$ the BHM and the continuous model predict the same transition curve.
There is a certain discrepancy between the two models at lower ratios, as for instance one gets $|a_{\rm 1D}|/a_0 = 0.402$ for $N=12$ and $|a_{\rm 1D}|/a_0 = (4/3)(J/U)_{\rm c}=0.373$ in the thermodynamic limit for $V_0=0$ in the BHM.
However, deviations are not as dramatic as for other quantities, for instance the one-body density matrix which is discussed later.
The SG model should be valid for shallow lattices, but it is not clear, within the model, up to which maximum value of $V_0$ it works.
It is kind of a surprise that the SG model coincides with the DMC result only at $V_0=0$, deviating from it for any finite value of $V_0$.
There is an overall good agreement with the experimental position of the phase transition~\cite{Haller10}.
In the region of shallow lattices, the amplitude modulation measurements are compatible with the DMC results, while the transport measurement at the weakest lattice agrees better with the SG model.

\begin{figure}[t!]
\begin{center}
\includegraphics[width=0.8\columnwidth, angle=0]{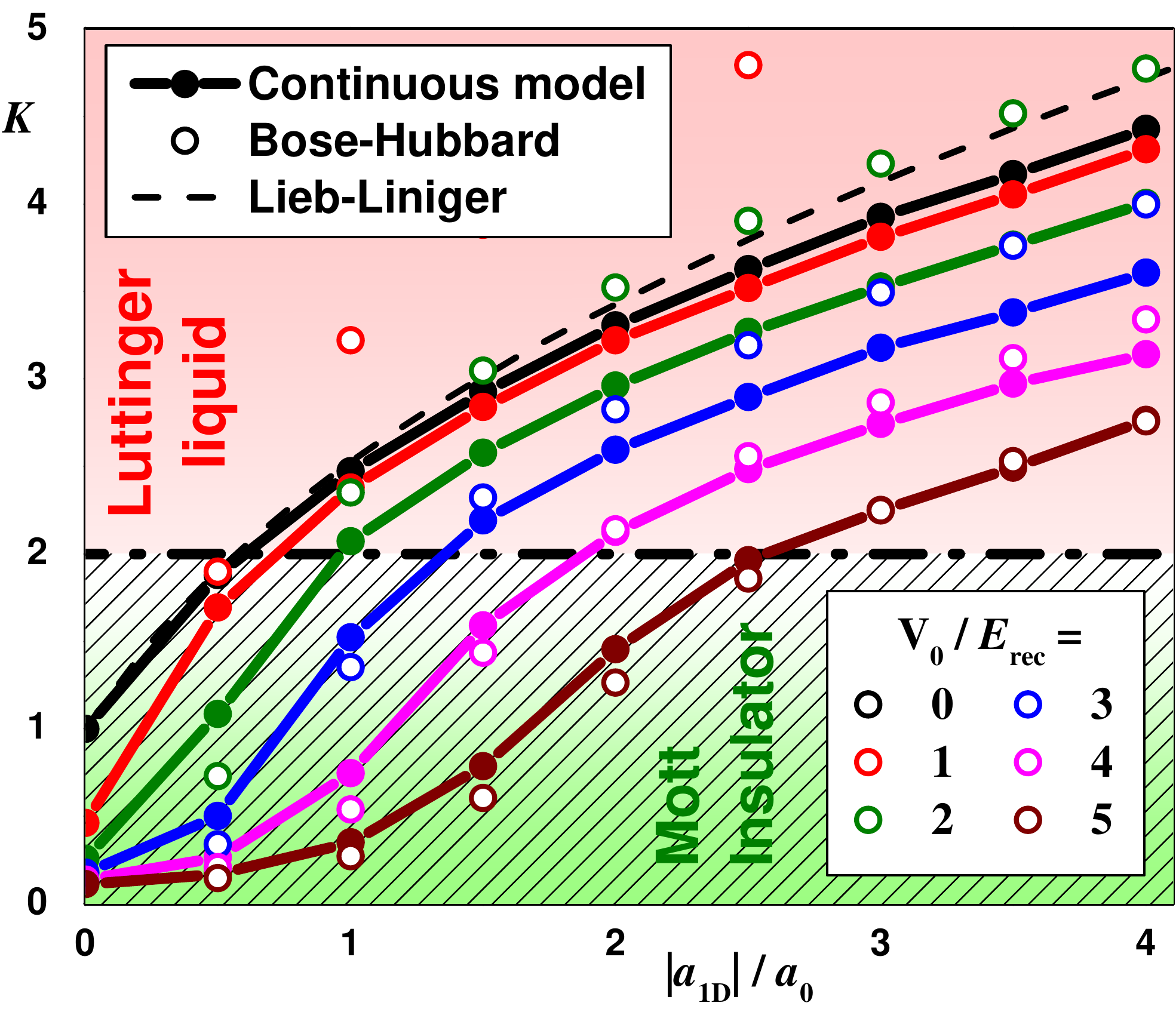}
\caption{(Color online) Parameter $K = 2S(k_{\rm min}) k_{\rm F}/k_{\rm min}$ calculated at the first point, $k_{\rm min} = 2\pi/(La_0)$,
of the static structure factor
for $N=12$ particles as a function of interaction strength $|a_{\rm 1D}|/a$ for lattices of different heights $V_0 = 0,1,2,3,4,5 E_{\rm rec}$
(top to bottom curves).
Solid symbols connected with solid lines, DMC results;
open symbols, BHM;
dashed line, Lieb--Liniger thermodynamic result for $V_0 = 0$.
The $K=2$ short-dashed line separates the superfluid regime (above) from the insulating one (below).
Notice that $K$ can be identified with the Luttinger parameter in the region with $K>2$.}
\label{fig2}
\end{center}
\end{figure}

Figure~\ref{fig2} reports the Luttinger parameter $K$ of the continuous model as a function of $|a_{\rm 1D}|/a_0$ for a number of characteristic values of $V_0$.
The figure also shows the BHM prediction and the $V_0=0$ Lieb-Liniger limit.
Its knowledge is essential in order to use the Luttinger liquid (LL) theory, which provides a description of long-range and small-momentum correlation functions.
It is important to realize that the effective LL theory uses $K$ as an input, while a full quantum many-body problem needs to be solved in order to obtain the dependence of $K$ on the system parameters.
The condition $K=2$ provides the critical value of $|a_{\rm 1D}|/a_0$ corresponding to the superfluid-insulator transition.
For $V_0 = 0$ the line starts exactly at $K=1$ (for Tonks-Girardeau $a_{\rm 1D}=0$ gas) and increases with the scattering length.
In that case the DMC results are compatible with the Bethe {\it ansatz} solution of the Lieb-Liniger model in the thermodynamic limit, while the deviations at weak interactions can be attributed to finite-size corrections.
In the Mott insulator regime, the sound is absent resulting in vertical lines for $K<2$ in the thermodynamic limit\cite{Hu2009}.
Within the BHM, $K$ depends on the single parameter $J/U$, generating a series of curves scaled by the value of $a_{1D}$.
For shallow lattices BHM predictions lie above the Lieb-Liniger $V_0=0$ curve, which by itself can serve as a test of validity of the BHM.
A more precise boundary of applicability is obtained when compared with the DMC results.
We find good agreement for large $|a_{\rm 1D}|/a_0$ and $V_0$.
For the range reported in Fig.~\ref{fig2} agreement for $K$ is achieved for $V_0/E_{\rm rec}\gtrsim 4$.

\begin{figure}[t!]
\includegraphics[width=0.8\columnwidth, angle=0]{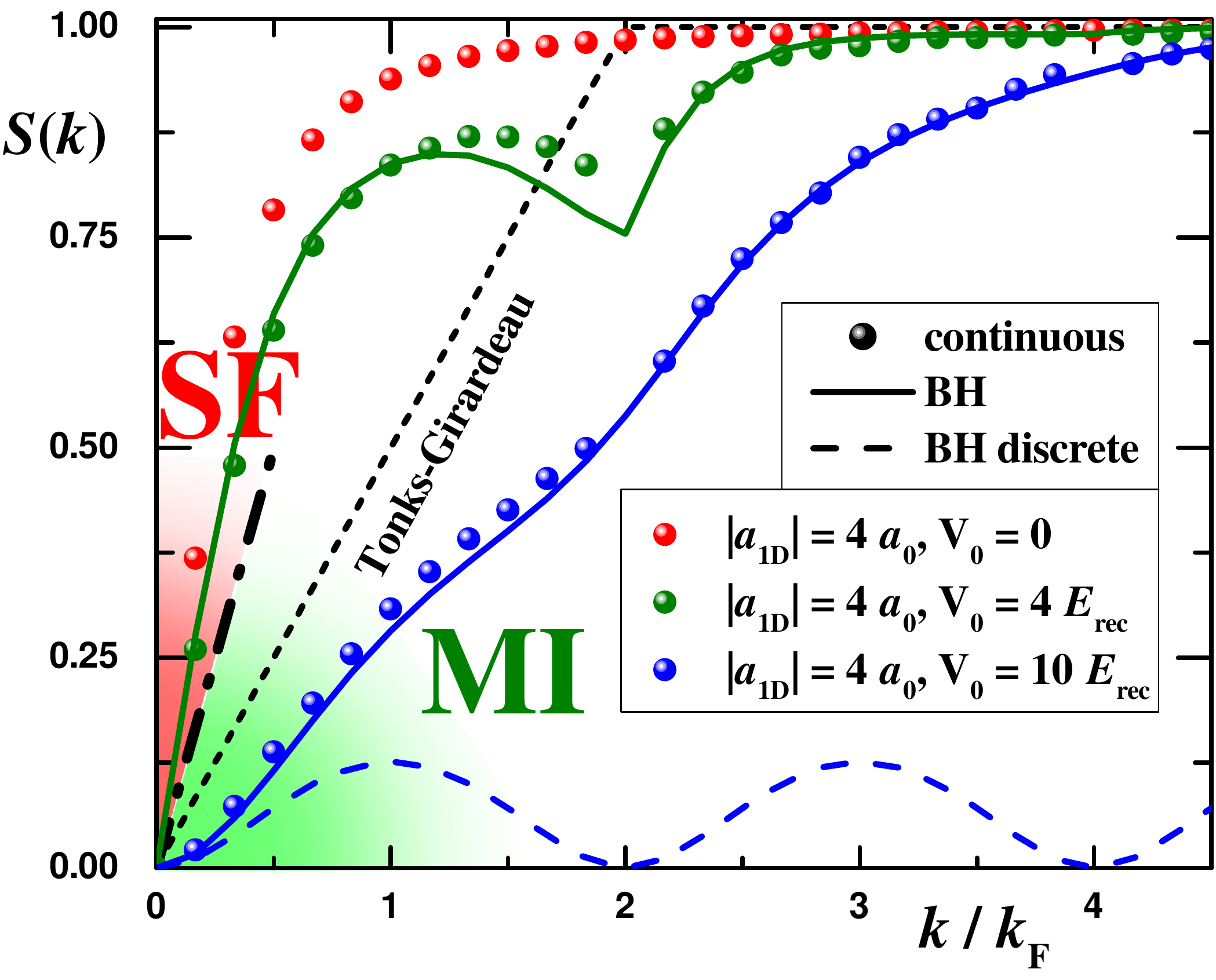}
\begin{center}
\caption{(Color online) Static structure factor $S(k)$.
Solid symbols, continuous model;
solid lines, $S_{\rm BH}(k)$ defined by~(\ref{S0lBB});
dashed line, $S_0(k)$ defined by~(\ref{S0discrete});
dash-dotted line, $S(k) = |k|/k_{\rm F}$ corresponds to a linear slope with the critical value of the Luttinger parameter $K=2$ and separates superfluid (SF) and Mott--insulator (MI) phases.
The following parameters are used:
short-dashed line, Tonks--Girardeau gas (i.e. $V_0 = 0$ and $a_{\rm 1D} = 0$);
continuous and Bose--Hubbard models (from top to bottom)
$V_0  / E_{\rm rec} = 0; 4; 8$ and $|a_{\rm 1D}|= 4a_0$.
}
\label{fig3}
\end{center}
\end{figure}

The static structure factor~\cite{Sengstock2010,Kozuma1999,Stenger1999,Stamper1999} $S(k)$  is defined as
$S(k)=\frac{1}{N} \langle\Delta{\tilde\rho}(k)\Delta{\tilde\rho}(-k)\rangle$
where $\Delta{\tilde\rho}(k)$ is the Fourier transform of the density-fluctuation operator (following the notations from Ref.~\cite{KrutitskyROPP}).
From the solution of the BHM it can be obtained as~\cite{KrutitskyROPP}
\begin{eqnarray}
\label{S0lBB}
S(k)
\approx
S_{\rm BH}(k)
=
1
+
G_0^2(k)
\left[
    S_0(k)
    -1
\right]
\;,
\end{eqnarray}
where
$G_0(k)
=
\int_0^{La_0}
d{x}
\left|
    W_\ell(x)
\right|^2
\exp
\left[
    -i k
    \left(
        x - x_\ell
    \right)
\right]
$  and
\begin{eqnarray}
\label{S0discrete}
S_0(k)
=
\frac{1}{N}
\!\!\sum_{l_1,l_2}
\left(
\langle
    \hat n_{l_1}
    \hat n_{l_2}
\rangle
\!-\!
\langle
    \hat n_{l_1}
\rangle
\langle
    \hat n_{l_2}
\rangle
\right)
\!\exp
\left[
    i k a_0
    \left(
        {l_2}-
        {l_1}
    \right)
\right]
\end{eqnarray}
is a discrete analogue of $S(k)$.
The typical behavior of $S(k)$ is shown in Fig.~\ref{fig3}.
In the Lieb-Liniger gas, corresponding to $V_0=0$, and large $a_{\rm 1D}$ the $S(k)$ is a featureless monotonous function typical for weakly interacting Bose gas ~\cite{AG03,AG06}.
The limit of $V_0=0$ and $a_{\rm 1D}=0$ corresponds to Tonks-Girardeau gas~\cite{Girardeau1960} which can be mapped to an ideal Fermi gas showing a kink at $k=2k_{\rm F}$, with the Fermi momentum $k_{\rm F}=\pi N/(La_0)$.
For a finite optical lattice, more features appear at momentum $k=2k_{\rm F}$, which corresponds to the border of the first Brillouin zone.

At low momenta, the static structure factor is well approximated by the Feynman relation:
\begin{equation}
\label{Sk-small}
S(k)
=
\frac{\hbar^2 k^2}{2m\epsilon(k)}
\;,\quad
\epsilon(k)
=
\sqrt
{
 \Delta^2
 +
 \left(
     \hbar c k
 \right)^2
}
\;,
\end{equation}
where $\Delta \geq 0$ is the energy gap.
Note that in units of $k/k_{\rm F}$ there is a critical value of the slope in $S(k)$ corresponding to $K=2$, see Fig.~\ref{fig3}.
If $S(k)$ lies above the critical line for small momentum, the gas is superfluid and $S(k)$ is linear for $k\to 0$, otherwise the system is insulating and $S(k)$ is quadratic for $k\to 0$.

\begin{figure}[t!]
\begin{center}
\includegraphics[width=1\columnwidth, angle=0]{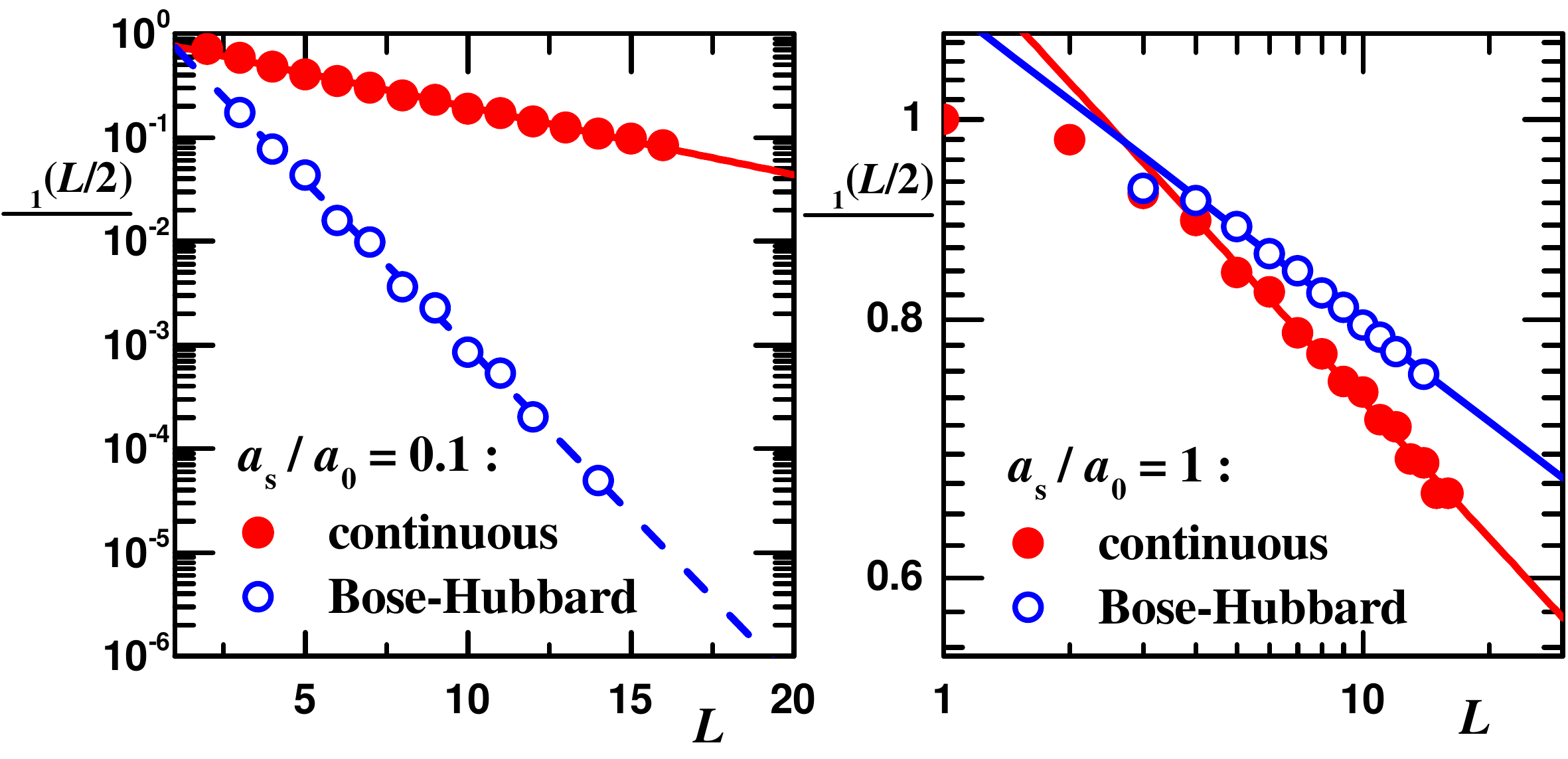}
\caption{(Color online) OBDM $\rho_1(L/2)$ obtained from the pure DMC estimator for $V_0 = 1 E_{\rm rec}$ and two characteristic values of the interaction strength, $|a_{\rm 1D}|/a_0 = 0.1$
(insulating phase, left panel) and $|a_{\rm 1D}|/a_0 = 1$ (superfluid phase, right panel) for different system sizes $L$.
Solid symbols, continuous model; open symbols, BHM.
For $|a_{\rm 1D}|/a_0=0.1$ one observes an exponential decay (insulating phase),
while for $|a_{\rm 1D}|/a_0=1$ it is much better fit by a power law form
(superfluid phase).}
\label{fig4}
\end{center}
\end{figure}

The coherence properties differ significantly in the insulating and the superfluid phases.
Commonly, the superfluidity is associated with the presence of a Bose--Einstein condensate, where the condensate wave function is the order parameter of the superfluid phase, manifested by the off-diagonal long-range order (ODLRO) in the one-body density matrix (OBDM) $\rho_1(r)$.
A finite value of the condensate fraction, $\rho_1(r\to\infty)/\rho\neq 0$, was used to localize the superfluid--Mott-insulator phase transition in three dimensions\cite{Pilati12}.
However, in one dimension quantum fluctuations destroy ODLRO, even at zero temperature~\cite{Hohenberg67}, and $\rho_1(r)$ always decays to zero.
The LL theory predicts a slow power-law decay in the superfluid phase~\cite{Haldane81}, in contrast with the fast exponential decay in the insulating phase.
Figure~\ref{fig4} shows the OBDM $\rho_1(r)$ calculated in two different phases at the largest possible length $r=L/2$ in a box of size $L$.
The solid lines show fits to the numerical data.
As it can be seen, the large distance behavior of $\rho_1$ is different in both cases, as for small values of $|a_{\rm 1D}|/a_0$ the OBDM presents an exponential decay, while in the opposite limit it is better reproduced by a power law.
The comparison with the BHM shows qualitative agreement in the form of the decay, while quantitatively the description of the discrete model can be quite off, especially for strong interactions.

\begin{figure}[t!]
\begin{center}
\includegraphics[width=0.8\columnwidth, angle=0]{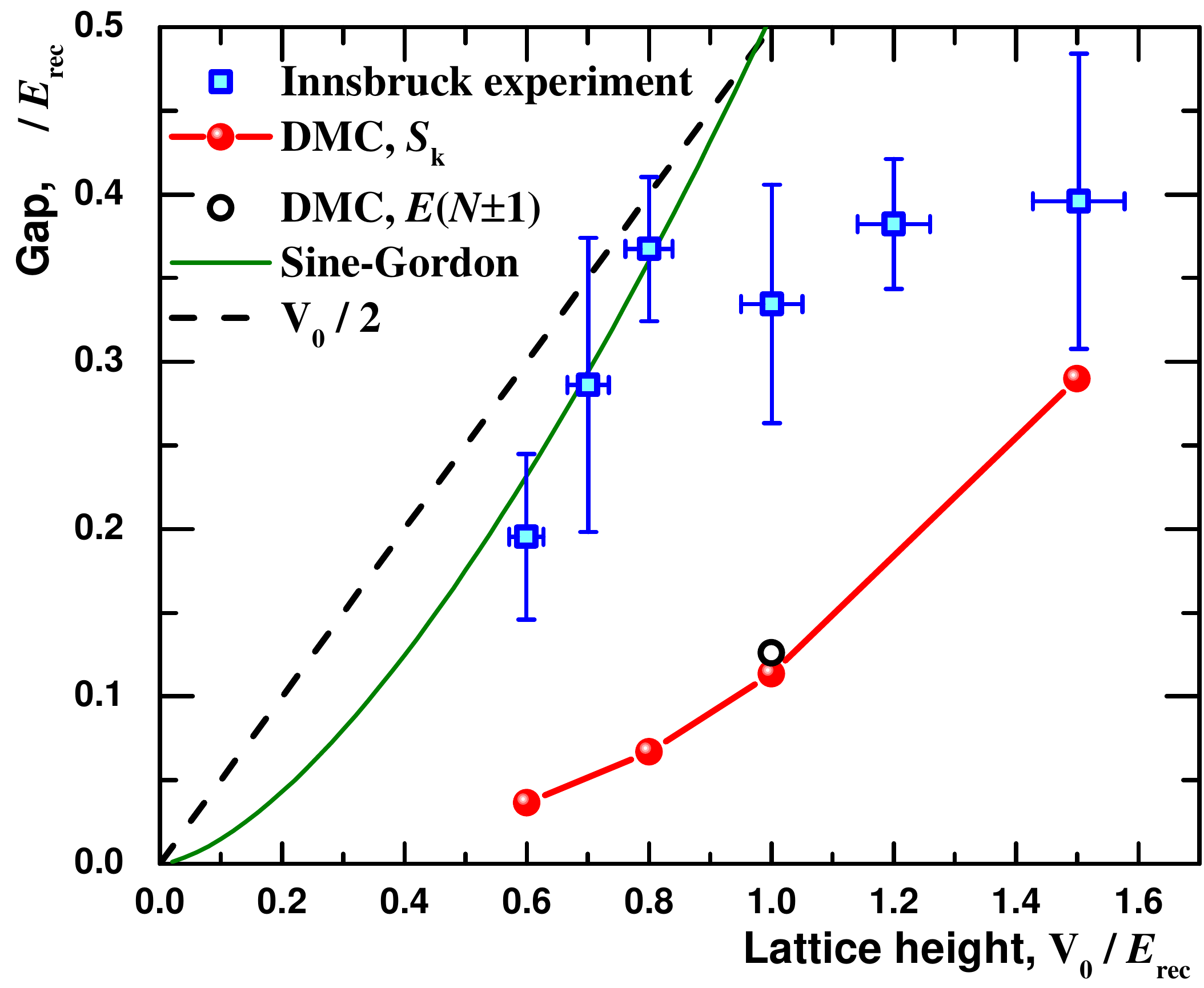}
\caption
{(Color online)
Energy gap.
Squares with error bars,
experimental data from Ref.~\cite{Haller10};
filled circles,
DMC prediction obtained from the fit to $S(k\to 0)$;
open circle,
DMC prediction obtained according to Eq.~(\ref{Eq:gap:E});
solid line,
sine--Gordon model~\cite{Haller10};
dashed line,
upper bound given by $V_0/2$.
$a_{\rm 1D}/a_0=0.18182$.
}
\label{fig5}
\end{center}
\end{figure}

The energy gap can be considered as an order parameter describing the insulating phase.
Figure~\ref{fig5} reports the gap calculated with different methods.
In the first one, the (charged) gap is evaluated from the ground-state energies calculated for $N$, $N+1$ and $N-1$ particles on $L=N$ lattice sites, according to the expression
\begin{equation}
\Delta_{\rm c} = E_{N+1} - 2 E_N + E_{N-1}
\;,
\label{Eq:gap:E}
\end{equation}
which corresponds to the difference of chemical potentials between the $N+1$ and the $N$ particle systems, respectively.
Alternatively, an upper bound for the gap is obtained from the Feynman relation~(\ref{Sk-small}).
We obtain $\Delta$ and $c$ by numerical fitting of the DMC data for $S(k)$.
The experimental data for the gap is taken from Ref.~\cite{Haller10}.
We observe a remarkable divergence between the {\em exact} DMC results, which are consistent among themselves using two different criteria, the SG model and the experiment.
Still, the experimental points are in better agreement with the SG model than with the DMC calculation.
This poses a question if the modulation spectroscopy method used in Ref.~\cite{Haller10} is precise for measuring the value of the gap in shallow lattices.
The gap $\Delta_{\rm c}$ obtained from the BHM (not shown) grows from $0.9$ for $V_0=0.5$ to $1.5$ for $V_0=1.5$ (in the units of $E_{\rm rec}$) for the parameters of Fig.~\ref{fig5} and lies above the $V_0/2$ line.
In this regime, $\Delta_{\rm c}$ is larger than the energy gap between the Bloch bands and the BHM is not valid.

To conclude, we established the zero-temperature phase diagram of a one-dimensional Bose gas in an optical lattice, determining the superfluid -- Mott insulator transition line.
We analyzed and compared the properties of a continuous Hamiltonian (using DMC method) with that of the discrete Bose--Hubbard model (solved via exact diagonalization).
We established the previously unknown regions of applicability of the approximate Bose--Hubbard and sine--Gordon models, and found that the sine--Gordon model fails to describe the regime of a shallow lattice for any finite lattice strength.
This poses a natural question if it is possible reconcile the discrepancy by improving the sine--Gordon model.
In general, Bose--Hubbard model is valid for high optical lattices with weak interactions, but the precise applicability of this description depends on the quantity of interest.
The dependence of the Luttinger parameter $K$ on the height of the lattice and the strength of the interaction is reported.
We also showed that the one-body correlations decays to zero following a slow power law in the superfluid phase, and exponentially in the Mott insulator phase.
We compared our results with the experiment of Ref.~\cite{Haller10}, and found an overall good agreement for the phase diagram.
Instead, we saw a discrepancy in the value of the excitation gap.
Importantly, our results help to understand the experiments with one-dimensional gases beyond the Bose--Hubbard approximation.

\begin{acknowledgments}
We thank Marcello Dalmonte, Eugene Delmer, Stefano Giorgini, Leonid Glazman, Bertrand Halperin, Lev Pitaevskii, Guido Pupillo and Klaus Sengstock for discussions. 
ML acknowledges ERC AdG OSYRIS, EU IP SIQS, EU STREP EQuaM, EU FET-Proactive QUIC, Spanish MINECO Project FOQUS (Grant No. FIS2013-46768-P), "Severo Ochoa" Programme (SEV-2015-0522), and the Generalitat de Catalunya Project 2014 SGR 874. 
The Barcelona Supercomputing Center (The Spanish National Supercomputing Center -- Centro Nacional de Supercomputaci\'on) is acknowledged for the provided computational facilities.

\paragraph*{Note added.---}
After the present work was completed and submitted for publication, a new experimental study~\cite{Boeris15} of the phase diagram by the LENS group in Florence appeared, in particular analyzing shallow lattices where we find discrepancy with the sine-Gordon model and the transport measurements of Ref.~\cite{Haller10}. 
The experimental measurements and the path integral Monte Carlo calculations of Ref.~\cite{Boeris15} agree with our predictions.
\end{acknowledgments}


\end{document}